\documentclass[adp,fleqn]{w-art}
\usepackage{times}
\usepackage{w-thm}
\usepackage[]{graphicx}
\chardef\bslash=`\\ % p. 424, TeXbook

\def\beq{\begin{equation}}
\def\eeq{\end{equation}}
\def\bea{\begin{eqnarray}}
\def\eea{\end{eqnarray}}
\def\nnu{\nonumber}
\def\tst{\textstyle}

\def\fno#1{Fig.~\ref{#1}}

\def\eno#1{Eq.~(\ref{#1})}

\def\Sno#1{Sec.~\ref{#1}}

\def\by{\over}
\def\gtwid{\mathrel{\raise.3ex\hbox{$>$\kern-.75em\lower1ex\hbox{$\sim$}}}}
\def\ltwid{\mathrel{\raise.3ex\hbox{$<$\kern-.75em\lower1ex\hbox{$\sim$}}}}

\def\al{\alpha}

\def\dta{\delta}
\def\eps{\epsilon}
\def\tta{\theta}

\def\sig{\sigma}
\def\om{\omega}

\def\Gam{\Gamma}
\def\Dta{\Delta}

\def\ptl{\partial}

\def\hf{{1\over2}}
\def\tshf{{\tst\hf}}

\def\part#1#2{{\ptl#1 \by \ptl#2}}

\def\ham{{\cal H}}
\def\ket#1{|#1\rangle}
\def\bra#1{\langle#1|}

\def\mel#1#2#3{\langle#1|#2|#3\rangle}

\def\bev{{\bf e}}

\def\bq{{\bf q}}

\def\bu{{\bf u}}
\def\bx{{\bf x}}

\def\bH{{\bf H}}

\def\bS{{\bf S}}

\def\xhat{{\bf{\hat x}}}

\def\zhat{{\bf{\hat z}}}

\def\itz{{\it z\ }}

\def\Fe8{Fe$_8$}
\def\Mn12{Mn$_{12}$}

\begin{document}
%%
%%  Most of the following commands will be completed by the publisher.
%%
%%  The copyrightyear is defined in the .clo file as the first argument
%%  of the copyrightinfo command. If the copyrightyear differs from that
%%  value it might be adjusted by the following definition:
%%
%% \renewcommand{\copyrightyear}{2002}% uncomment to change the copyrightyear.
%%
\DOIsuffix{theDOIsuffix}
%%
%% issueinfo for header and copyright line
%%
\Volume{12}
\Issue{1}
\Copyrightissue{01}
\Month{01}
\Year{2003}
%%
%%  First and last pagenumber of the article. If the option
%%  'autolastpage' is set (default) the second argument may be left empty.
\pagespan{1}{}
%%
%%    Dates will be filled in by the publisher. The 'reviseddate' and
%%    'dateposted' (Published online) entry may be left empty.
\Receiveddate{15 November 1900}
%%
%% \Reviseddate{30 November 1900}
%%
\Accepteddate{2 December 1900}
%% \Dateposted{3 December 1900}
%%
%%

%%
%%\subjclass[pacs]{04A25, ...} % up to three, separated by commas
%% NOTE: PACS will NOT be needed anymore after January 1, 2010! 
%%

%% \pretitle{Editor's Choice}

%% We have a short and a long form for the title. The short form
%% (optional argument) goes into the running head.

\title[Phonoemissive Spin Tunneling]{Phonoemissive Spin Tunneling in Molecular Nanomagnets}

%% Please do not enter footnotes or \inst{}-notes into the optional
%% argument of the author command. The optional argument will go into
%% the header. If there is only one address the marker \inst{x} may be
%% omitted.

%%   Information for the first author.
\author[Anupam Garg]{Anupam Garg\footnote{Corresponding
     author \quad E-mail: {agarg@northwestern.edu}}}%\inst{1}} 
%\address[\inst{1}]{Department of Physics and Astronomy, Northwestern University,
\address{Department of Physics and Astronomy, Northwestern University,
Evanston, Illinois 60208, USA}

%%  \dedicatory{This is a dedicatory.}
%%
\begin{abstract}
A new mechanism is proposed for the magnetization reversal of molecular nanomagnets such as \Fe8.
In this process the spin tunnels from the lowest state near one easy direction to the first excited
state near the opposite easy direction, and subsequently decays to the second easy direction with
the emission of a phonon, or it first emits a phonon and then tunnels to the final state.
This mechanism is the simplest imaginable one that allows magnetization
relaxation in the presence of a longitudinal magnetic field that is so large that the nuclear spin
environment cannot absorb the energy required for energy conservation to hold. It is proposed
as a way of understanding both magnetization realaxation and Landau-Zener-St\"uckelberg experiments.
The requisite Fermi golden rule rate, and the spin-flip rates are calculated, and it is found that
these rates are much too low by several orders of magnitude. Thus the understanding of magnetic
relaxation in the experiments remains an open question.
\end{abstract}
\maketitle                   

%% If there is not enough space inside the running head
%% for all authors including the title you may provide
%% the leftmark in one of the following three forms:

%% \renewcommand{\leftmark}
%% {F.\ Author: Short Title}

%% \renewcommand{\leftmark}
%% {F.\ Author and S.\ Author: Short Title}

%% \renewcommand{\leftmark}
%% {F.\ Author et al.: Short Title}

%% \tableofcontents  % Produces the table of contents.

%zzzypt

%\twocolumn[
%\hsize\textwidth\columnwidth\hsize\csname@twocolumnfalse\endcsname
%
\date{\today}

%\pacs{75.50.Xx, 75.60.Jk, 76.20.+q, 75.45.+j}

\section{Introduction and Background}
\label{intro}

Over the last fifteen years or so, molecular nanomagnets (also known as single-molecule
magnets, or molecular magnets) have provided us with an entirely new class of magnetic systems,
showing many novel phenomena not seen previously~\cite{gsvbook}.
Among the most dramatic of these is the
observation of {\it gap oscillations\/} wherein the tunnel splitting between the two lowest
energy states on opposite sides of an energy barrier oscillates as a function of a static
magnetic field applied along the hard axis of the molecule~\cite{ww1}.
Although many aspects of this phenomenon can be understood by considering the idealized
problem of an isolated molecule, a full understanding of the experimental procedures and
results presents several complexities and puzzles. It is the purpose of this paper to
address one of these puzzles. To keep the discussion focussed and free of lapidary
generalizations,  we will consider the example of the \Fe8
molecule in this paper, but the central ideas are applicable more broadly.

To understand the basic phenomenon at issue, let us first consider one \Fe8 molecule in the solid,
and ignore its interaction with other molecules and any other environmental degree of freedom.
The molecule has spin 10 in its ground manifold. At low temperatures only the spin degree
of freedom has any life in it, and all others are frozen. This degree of freedom is
governed by an anisotropy Hamiltonian
\beq
\ham_{\rm s} = k_1 S_x^2 + k_2 S_y^2 +\ham_4 -g\mu_B \bS\cdot\bH,
                       \label{hams}
\eeq
where $\bS$ is a dimensionless spin operator (of magnitude $S$, equal to 10 for \Fe8),
$k_1 > k_2 > 0$ are anisotropy coefficients, $g$ is a g-factor (equal to 2 for \Fe8),
and $\bH$ is an
external magnetic field. The term $\ham_4$ is of fourth order in the components of
$\bS$, and while it is responsible for surprising and important features in the
tunneling spectrum, we do not show it explicitly because we will not need to discuss
those aspects in this paper.

When $\bH = 0$, \eno{hams} has two ground states corresponding (classically speaking) to
$\bS\|\pm \zhat$, or $m = \pm 10$, where $m$
is the $S_z$ or Zeeman quantum number. These two states are degenerate, and mixed by tunneling,
as shown in \fno{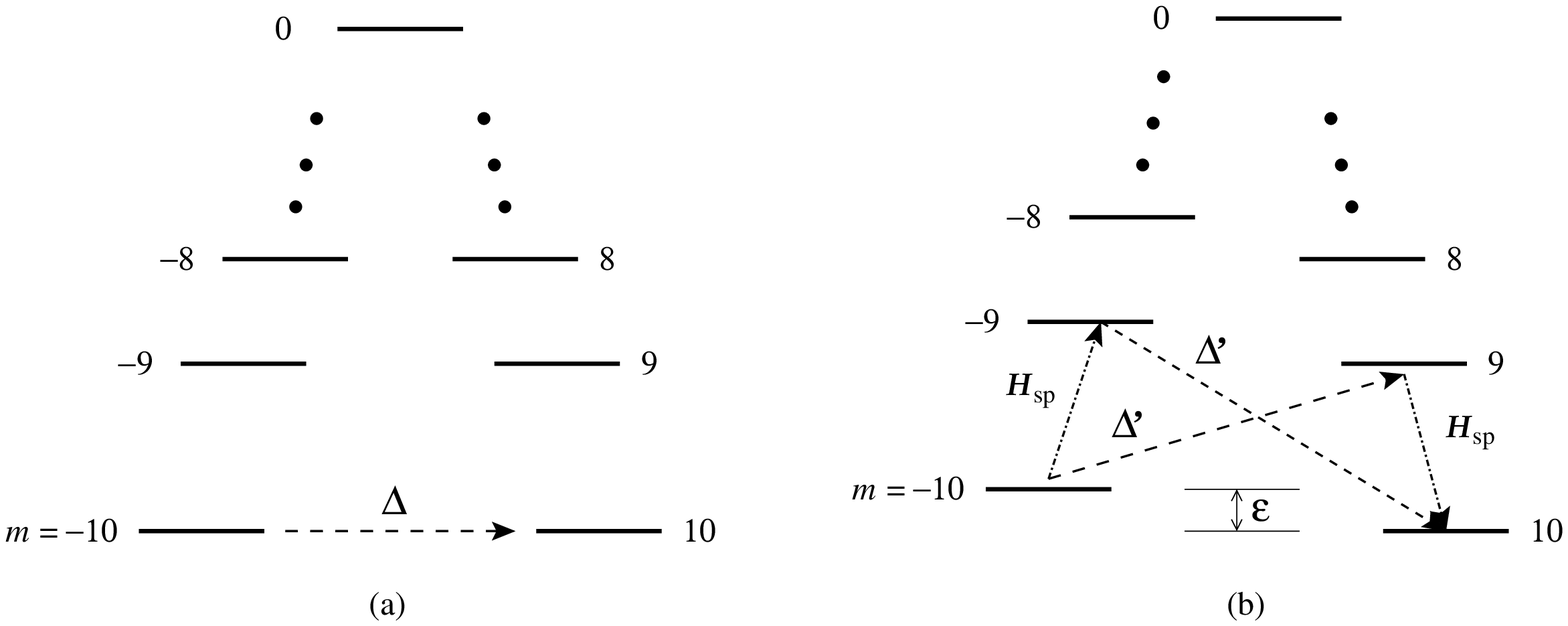}.
\begin{vchfigure}[htb]
  \includegraphics[width=.9\textwidth]{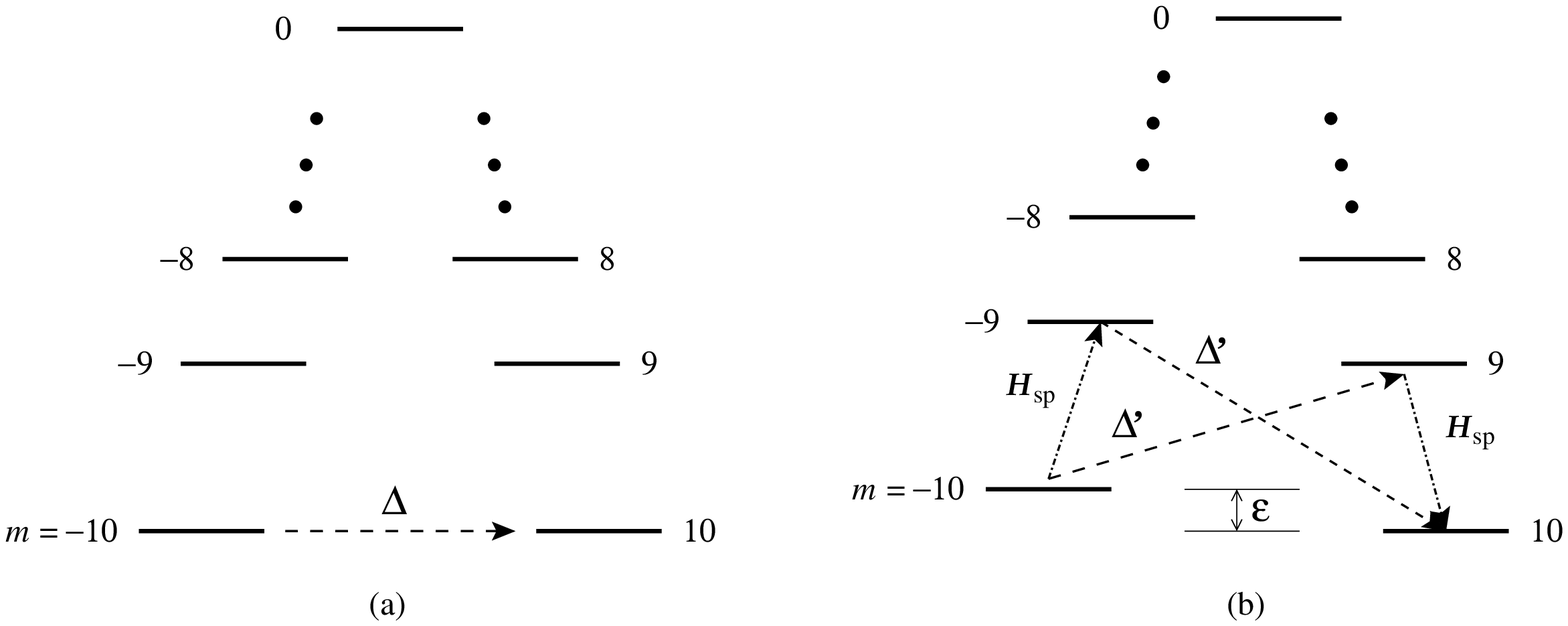}
\vchcaption{
Schematic energy level diagram of \Fe8 showing (a) the basic tunneling process between ground
levels, and (b) the phonoemissive process, in which the molecule tunnels from $m=-10$ to an
intermediate virtual state $m=9$, and then makes a transition to $m=10$ with the emission of a
phonon (as shown by the dot-dashed line). Or, the phonon can be emiited first accompanying
a transition to the $m=-9$ state, followed by tunneling to the $m=10$ state.
This process appears to be the simplest way in which
the magnetization can relax when the bias $\eps$ is much greater than $W$, the effective width
of the levels induced by the nuclear spin environment. Direct tunneling in this case would not
conserve energy, and the nuclear spins are incapable of absorbing an energy much greater than
$W$.
}
\label{elevels.eps}
\end{vchfigure}
If we now turn on $\bH\| \xhat$, the new classical ground states are still degenerate, but
the tunnel splitting between them ($\Dta_{-10,10} \equiv \Dta$)
does not increase monotonically with $H_x$. Instead it
oscillates as shown in \fno{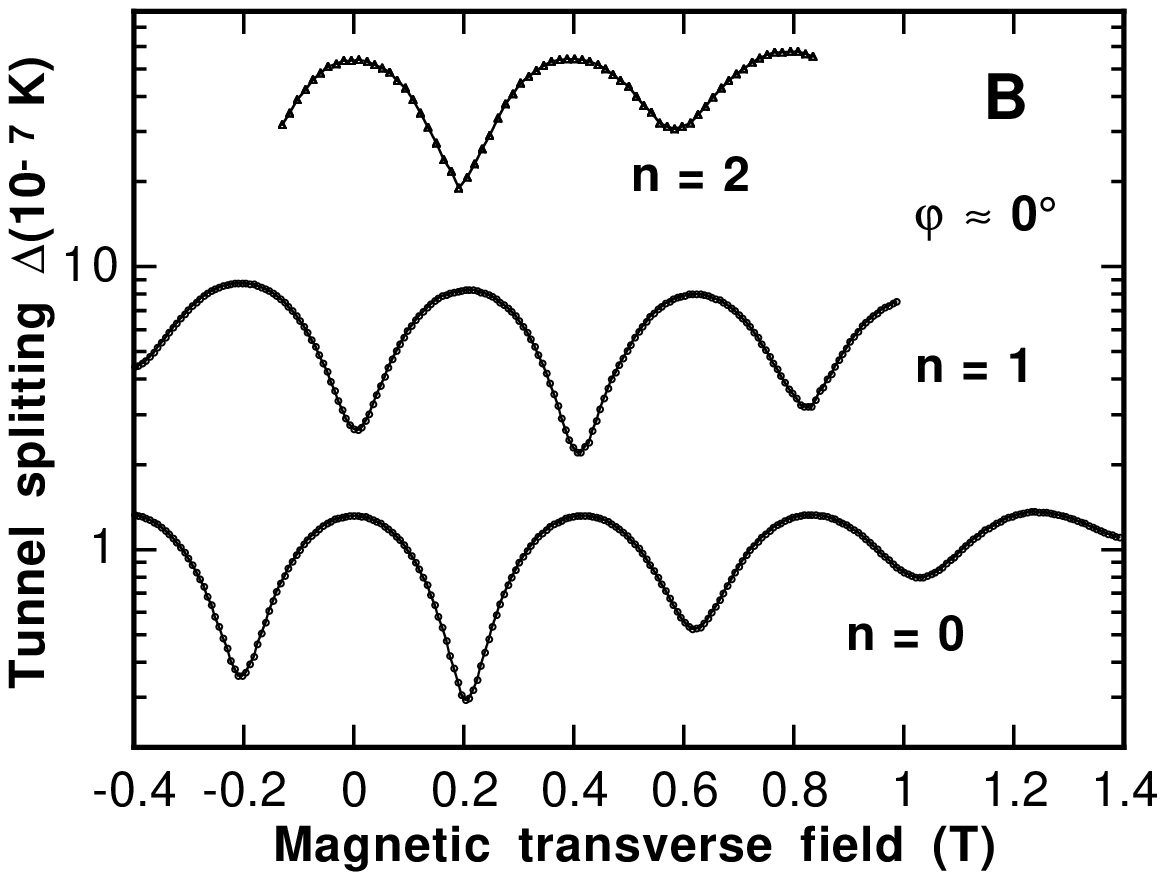}.
This oscillation is best understood in terms of instanton~\cite{loss,vdh,ag93},
but readers unfamiliar with instantons may regard it as an outcome of direct diagnolization
of the Hamiltonian (\ref{hams}). Similarly, if a magnetic field is applied along the \itz
axis so as to bring the $m=-10$ and $m=9$ states into resonance, the system will be able to
tunnel between these states. This tunnel splitting ($\Dta_{-10,9} \equiv \Dta'$) also oscillates
as a function of an additional field $H_x$ along the hard axis. The same holds for tunneling
between $m=-10$ and $m=8$, and other pairs of states. By symmetry, if the \itz axis field is
such to bring the $m=-9$ and $m=10$ states into resonance, the splitting $\Dta_{-9,10}$ will be
the same as $\Dta_{-10,9}$.

\begin{figure}[htb]
\includegraphics[width=.45\textwidth]{gap_oscs.ps}~a)
\hfil
\includegraphics[width=.45\textwidth]{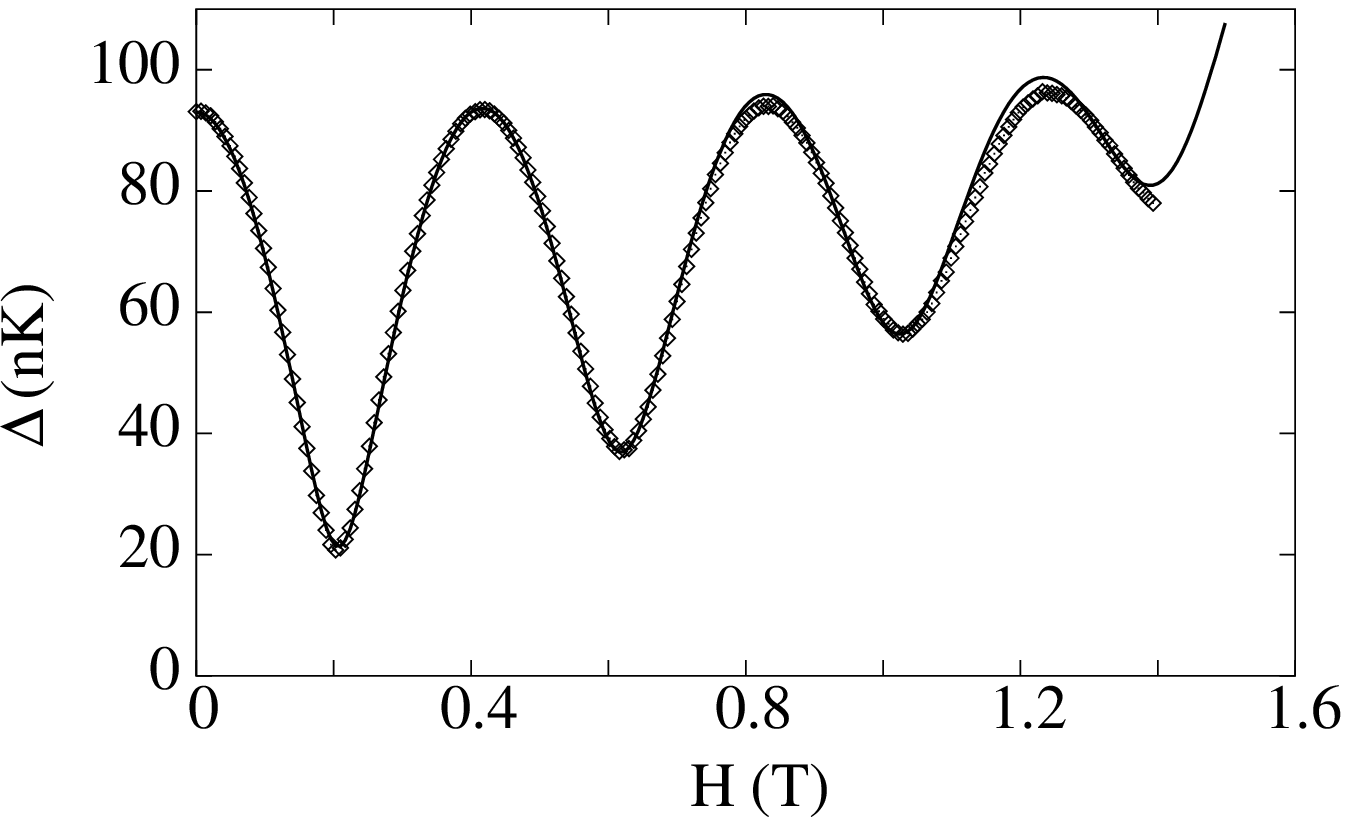}~b)
\caption{The gap oscillations in \Fe8. Part (a), kindly supplied by Dr.~Wernsdorfer, shows
data from Ref.~\cite{ww1}. The curve labeled $n=0$ is the $-10\to10$ tunneling, that labeled
$n=1$ is the $-10\to 9$ tunneling, and so on. Part (b) shows a fit to data for the tunnel
splitting between the lowest pair of levels, taken from Ref.~\cite{kg07}.}
\label{gap_oscs.ps}
\end{figure}

(To avoid misunderstanding, we note that when we speak of the state with $m=-10$, say, we do not literally
mean the eigenstate of $S_z$ with eigenvalue $-10$. Rather, we mean the eigenstate of $\ham_{\rm s}$ (ignoring
tunneling) that maps on to the $S_z = -10$ state in the following way. We imagine 
subtracting a zeroth order Hamiltoian $(k_1 +k_2)(S^2_x + S_y^2)/2$ from \eno{hams}, and treating
the result as a perturbation. The state that we are talking about is then the one that would develop
from the $S_z = -10$ state by low order perturbation theory. It would perhaps be better to use a notation
such as $m^*$ for the perturbed state, but as long as this qualification is understood, there is little
benefit from doing so.)

If an \Fe8 molecule really did not interact with the rest of the world, the tunneling described
above would lead to coherent flip-flop as in the inversion resonance of ammonia. No such flip-flop is seen,
and indeed that is to be expected. Any environmental degre of freedom which couples to the
magnetic moment of the molecule will tend to suppress quantum coherence, and for resonance between
states with such a large difference in their magnetic moments, one would expect that all vestiges
of coherence are destroyed. This is indeed so, and one finds that because of the nuclear spin
environment, transitions between the $\pm S$ states are totally incoherent, and one finds a
transition probability per unit time given by~\cite{pro98ab,avag09}
\beq
\Gam = {\sqrt{2\pi} \by 4} {\Dta^2 \by W}
          \exp -\left({\eps^2\by 2W^2} \right).  \label{p_flip}
\eeq
Here, $W \simeq 10 E_{dn}$, where $E_{dn}$ is the energy of dipole-dipole interaction
between the molecular electronic
spin and the nuclear spins of nearby nonmagnetic atoms such as N and H which
are always present in the molecules studied, and $\eps$ is the {\it bias\/}, or the energy of the
$m = -S$ state relative to the $m=S$ state. This bias can arise from an externally
applied field along the \itz axis, or from the dipole field created by other molecules in the
sample. Indeed, it is found that this dipole field is of order 100 Oe~\cite{ohm98,ww4},
so the bias is of order $0.1$\,K. By comparison, $E_{dn} \sim 1$\,mK.

Given that $\eps \gg W$ for most molecules in any solid sample, the spins of most molecules are
frozen, in that they do not even undergo incoherent tunneling. To overcome this problem, the authors of
Ref.~\cite{ww1} use the ingenious idea of sweeping through the resonance by applying a time-dependent
longitudinal (\itz axis) magnetic field as shown in \fno{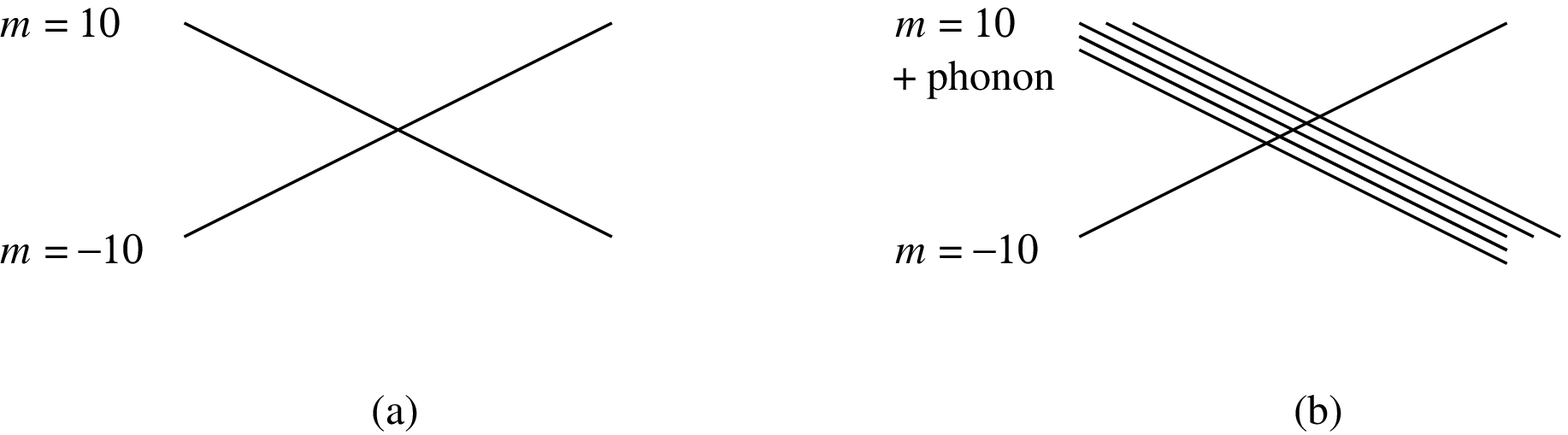}. This is the Landau-Zener-St\"uckelberg 
(LZS) protocol~\cite{lzs32}. The molecule flips from $m=-10$ to $m=10$ (or vice versa) with a probability 
\beq
p_{\rm LZS} = 1 - \exp\left(-{\pi \Dta^2 \by 2 \hbar |\dot\eps|}\right) \label{plzs}
\eeq
for every sweep through the crossing. Here $\dot\eps$ is the rate at which the bias changes. For
transitions between $m=\pm S$, $|\dot\eps| = 4\mu_B S|dH_z/dt|$, since $g=2$. It is found that for
${\dot H}_z$ between $\sim 3$\,mT/s and 1\,T/s, the LZS formula is obeyed, allowing one to extract
$\Dta$ from a measurment of $p_{\rm LZS}$.

\begin{vchfigure}[htb]
  \includegraphics[width=.9\textwidth]{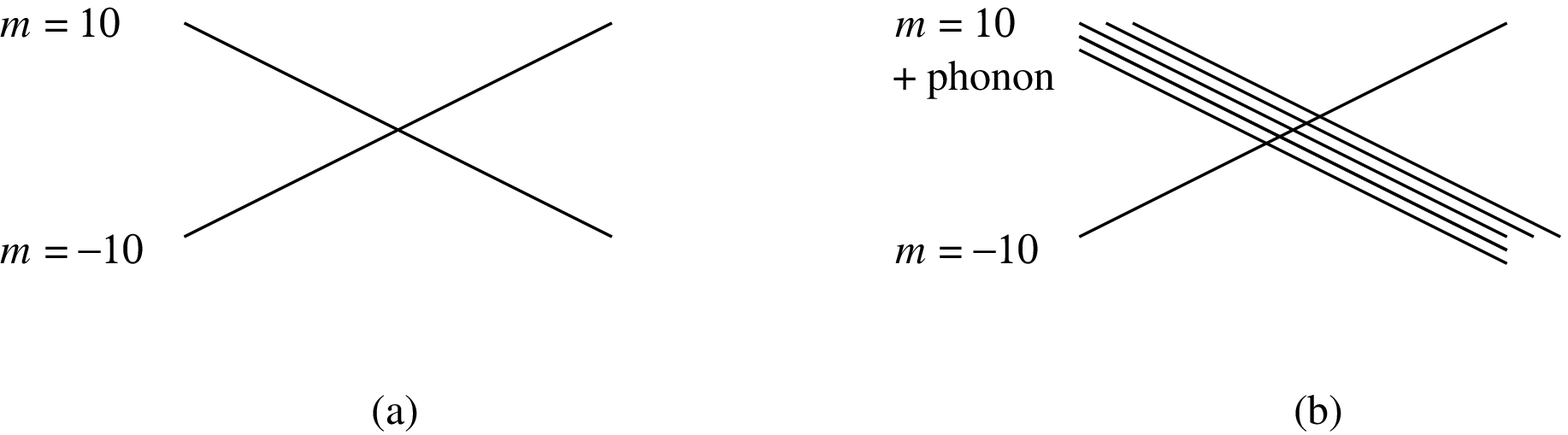}
\vchcaption{
The Landau-Zener-St\"uckelberg protocol. In part (a) we show the direct LZS mechanism, in which
the energy levels of the $m=\pm 10$ states are swept by applying a time-dependent field along
the \itz axis. In (b) we show the same process when the final state also contains a phonon.
}
\label{ecross.eps}
\end{vchfigure}

At this point it should be stated that the LZS formula (\ref{plzs}) only holds for coherent
transitions. For incoherent transitions, a different formula, due to Kayanuma~\cite{kaya84} is better:
\beq
p_{\rm K} = {1\by 2} \left[1 - \exp\left(-{\pi \Dta^2 \by\hbar |\dot\eps|}\right)\right]. \label{pk}
\eeq
Remarkably, for fast sweeps, i.e., when $|\dot\eps| \gg \Dta^2/\hbar$, both formulas agree:
\beq
p_{\rm LZS} = p_{\rm K} = {\pi \Dta^2 \by 2 \hbar |\dot\eps|},
             \quad (|\dot\eps| \gg \Dta^2/\hbar)
                           \label{pfast}
\eeq
Experimentally, the LZS formula starts to fail for $dH_z/dt \ltwid 3$\,mT/s (the Kayanuma formula
is not precisely obeyed either), but if one only employs it for fast sweeps, then the details of
the decoherence do not matter, and the extraction of $\Dta$ is reliable.

\section{The Phonoemssive Tunneling Process}
\label{pet}

We are now ready to describe the puzzle mentioned at the start of \Sno{intro}. Theoretically, at
the minima, $\Dta$ vanishes strictly. Experimentally, this is not so. In Ref.~\cite{kg07}, this
was explained by arguing that every molecule sat in a nonzero transverse magnetic
field created by the other molecules that acts in addition to the applied $H_x$.
Thus $\Dta \ne 0$ for any molecule, and the measured $\Dta$ should really be found by taking an average
over the inhomogeneous spread in $H_x$ of order 100 Oe. (Actually, the quantity that should be averaged
is $\Dta^2$.) In addition, the experimental data show that the inferred values of $\Dta$ at the minima
grow linearly with $H_x$. To explain this, it was assumed that there was a small degree of misalignment
of the magnetic axes of the molecules (of order $\sig_{\tta} \sim 1$--2$^{\circ}$) because of various types of defects
in the material, which is rather soft and organic. This means that there is a small nonzero field
$H_{\rm med} \simeq H_x \sig_{\tta}$ along the local {\it medium\/} axis of each molecule, and since 
for an isolated molecule $\Dta \propto H_{\rm med}$ in the vicinity of one of the minima, the measured
$\Dta$ would pick up a contribution linear in $H_x$.

As can be seen from \fno{gap_oscs.ps}(b), the fit between the theory based on these assumptions and the experimental
data is good; indeed it is rather too good. For while the first assumption---of an inhomogeneous transverse
dipolar field---is well justified and hardly an assumption, the second---of misalignment of the magnetic
axes---is somewhat {\it ad hoc\/}. We have therefore sought another way in which the spin could flip
for which the rate has an $H_x$ dependence. The mechanism we propose is shown in \fno{elevels.eps}(b). It
involves a virtual tunneling transition from the $m=-10$ to the $m=9$ state, followed by a transition
from the $m=9$ state to the $m=10$ state with the emission of a phonon. Alternatively, we could first make a
virtual transition to the $m=9$ state accompanied by the emission of a phonon, and then tunnel to the $m=10$
state. Either way, the final state is the same, i.e., the spin is in the $m=10$ state, and a phonon has been
emitted. We refer to this as phonoemissive tunneling. The reverse process, involving
the absorption of a phonon, will also take place if $k_B T \gtwid \eps$, for then phonons of energy
$\eps$ will be available to be absorbed. In this paper we work only at $T=0$, so only the emission
process takes place. The generalization to $T\ne 0$ is straightforward.

At this point it may be useful to clarify that the $-10\to 9$ (or $-9\to 10$) tunneling transition is possible even if
the levels are not in resonance. This transition would have very small probability if we were thinking of the
$m=9$ state as a true final state (not to mention that it would require some environmental degree of freedomn
to supply the energy necessary for energy conservation), but we are only exploiting $m=9$ as
a virtual intermediate state. For the virtual process, there is always a matrix element of the Hamiltonian,
or tunneling {\it amplitude\/}, equal to $-i\Dta'/2\hbar$ per unit time. Indeed, for \Fe8, $\Dta'$ as a
function of $H_x$ is maximal when $\Dta$ is minimal.

The process just described could also be germane to experimental situations other than LZS. Consider,
for example, magnetic relaxation without any swept field. If, for a given molecule, $\eps \gg W$ for
some reason, the rate (\ref{p_flip}) is very small, and phonoemisive tunneling may be important. Indeed,
we shall calculate the second order Fermi golden rule rate for this situation first, in
\Sno{fgr}, before considering the LZS protocol in \Sno{lzs}. The process could also be important for
ultraslow LZS sweeps, and for inverse LZS sweeps~\cite{werns05}.

The second step of phonoemissive tunneling is governed by the spin-phonon Hamiltonian,
\beq
\ham_{\rm sp} = D_{ijkl} u^{\rm op}_{ij} \{S_k, S_l\}. \label{hsp_full}
\eeq
Here, $u^{\rm op}_{ij}$ is the operator for the strain tensor, $D_{ijkl}$ is the magnetoelastic tensor,
$\{,\}$ denotes the anticommutator, and there is an implicit sum over the Cartesian indices
$i$, $j$, $k$, and $l$. The tensor $D_{ijkl}$ is not well known for \Fe8, so we shall simplify
\eno{hsp} to
\beq
\ham_{\rm sp} = D u^{\rm op}_{ij} \{S_i, S_j\}. \label{hsp}
\eeq
The quantity $D$ has dimensions of energy, and may be taken to be of the same order of magnitude
as the anisotropy coefficients $k_1$ and $k_2$. The strain tensor is given by
\beq
u^{\rm op}_{ij} = {1\by 2}\left(\part{u^{\rm op}_i}{x_j} + \part{u^{\rm op}_j}{x_i} \right),
\eeq
where $\bu^{\rm op}(\bx)$ is the displacement field operator given by
\beq
\bu^{\rm op}(\bx) = \sum_{\al} \sqrt{\hbar\by 2 MN \om_{\al}} (a_{\al} \bev_{\al} e^{i\bq_{\al}\cdot\bx} + {\rm h.c.}).
              \label{disp}
\eeq
Here, $a_{\al}$ and $a^{\dagger}_{\al}$ are the destruction and creation operators for phonons of
mode $\al$, $\bq_{\al}$, $\om_{\al}$, and $\bev_{\al}$ are the wavevector, frequency, and polarization
vector for this mode, $M$ is the mass of all atoms in a unit cell, and $N$ is the number of unit
cells in the crystal. In writing \eno{disp}, we have kept only long wavelength phonon modes, for
which relative motion of the atoms in a unit cell is negligible, and this is why the total unit
cell mass $M$ appears. These are the only modes relevant for low temperatures and low energy processes.
We can also take the polarization vectors to be real for them.

For the $m=9$ to $m=10$ transition, the only nonzero matrix elements arise from the spin operators
$\{S_x,S_z\}$ and $\{S_y,S_z\}$. We have
\bea
\mel{m=S}{\{S_x, S_z\}}{m=S-1} &=& (S - \tshf)\sqrt{2S} \equiv A, \nnu\\
\mel{m=S}{\{S_y, S_z\}}{m=S-1} &=& -i(S - \tshf)\sqrt{2S} = -iA.
         \label{defA}
\eea
It is now apparent that we could also take an $m=8$ intermediate state, since the operator 
$\{S_x, S_y\}$ would yield a non zero matrix element, and there is a nonzero amplitude to tunnel from
$m=-10$ to $m = 8$. We will not give formulas for this case explicitly, as the requisite modifications are
straightforward. Other intermediate states are much less important, since they would involve
multiphonon processes with far smaller probabilities.

\section{Fermi Golden Rule}
\label{fgr}

We now calculate the rate for phonoemissive tunnelling when there is no swept field. We denote the initial,
intermediate (or virtual), and final states by $\ket{i}$, $\ket{v}$, and $\ket{\al}$, where,
\bea
\ket{i} &=& \ket{m=-10, {\rm no\ phonons}}, \\
\ket{v} &=& \ket{m=9, {\rm no\ phonons}}, \label{ivf} \\
\ket{\al} &=& \ket{m=10, {\rm one\ phonon\ in\ mode\ }\al}.
\eea
The energies of these states are
\beq
E_i = \eps_{-10}, \quad E_v = \eps_9, \quad E_{\al} = \eps_{10} + \hbar\om_{\al},
\eeq
where $\eps_m$ is the energy of the molecular spin alone. Alternatively, the transition could take place through
a different virtual intermediate state,
\beq
\ket{v} = \ket{m=-9, {\rm one\ phonon\ in\ mode\ }\al}, \\
\eeq
in which case,
\beq
E_v = \eps_{-9} + \hbar\om_{\al}.
\eeq
The transition rate is given by the second order Fermi golden rule,
\beq
\Gam_{\rm pet} = {2\pi \by\hbar} \sum_{\al} |V^{(2)}_{\al i}|^2 \dta(E_i - E_{\al}),
                    \label{Gpet_def}
\eeq
where
\beq
V^{(2)}_{\al i} = \sum_v {\mel{\al}{\ham_{\rm sp}}{v} \mel{v}{\ham_{\rm s}}{i} \by E_i - E_v}.
      \label{V2}
\eeq
It should be noted that the sum over virtual states in \eno{V2} does not entail a sum over the phonon
modes; rather it is performed for a particular mode $\al$. The sum over phonon modes is performed
in \eno{Gpet_def}.

It is evident that because of symmetry, the two
contributions to \eno{V2} differ only in the energy denominators. We shall assume that the bias $\eps$ is small
in comparison with the energy difference
\beq
E_{\rm ex} \equiv \eps_9 - \eps_{10}.   \label{eex}
\eeq
In that case, both energy denominators may be replaced by $- E_{\rm ex}$. We need only do the calculation for the
intermediate state (\ref{ivf}), and double the answer to get $V^{(2)}$. Further, it is more accurate to calculate
the energy difference using the results for $\eps_m$ when $H_z = 0$.

The actual calculation is straightforward. Consider the intermediate state (\ref{ivf}).
For the matrix element of $\ham_{\rm s}$ we have
\beq
\mel{v}{\ham_{\rm s}}{i} = \Dta'/2.
\eeq
Next let us examine the matrix element of $\ham_{\rm sp}$. Consider the strain field $u^{\rm op}_{xz}$.
We have
\beq
u^{\rm op}_{xz} = {i\by 2} \sum_{\al} \sqrt{\hbar\by 2 MN \om_{\al}} (a_{\al} - a^{\dagger}_{\al})
                       (e_{\al,x} q_{\al,z} + e_{\al,z} q_{\al, x}).
              \label{uxz_op}
\eeq
When we take the matrix element of this operator, we are left with the c-number
\beq
u^{\al}_{xz} = - {i\by 2} \sqrt{\hbar\by 2 MN \om_{\al}}
                       (e_{\al,x} q_{\al,z} + e_{\al,z} q_{\al, x}).
              \label{uxz}
\eeq
Similarly, from $u^{\rm op}_{yz}$ we get $u^{\al}_{yz}$, which is the same expresssion with the index
$x$ replaced by $y$ everywhere.  The matrix elements of the spin parts of $\ham_{\rm sp}$ have already
been found in \eno{defA}. Hence, doubling the result as explained above to account for the two
intermediate states, we get
\beq
V^{(2)}_{\al i} = -{AD \Dta' \by E_{\rm ex}} (u^{\al}_{xz} - i u^{\al}_{yz}).
\eeq
Therefore,
\beq
\Gam_{\rm pet}
    = {2\pi\by \hbar} \left( {AD \Dta' \by E_{\rm ex}} \right)^2
        \sum_{\al} |(u^{\al}_{xz} - i u^{\al}_{yz})|^2
           \dta(\hbar\om_{\al} - E_{if}), 
                  \label{Gpet}
\eeq
where
\beq
E_{if} = \eps_{-10} - \eps_{10} = \eps.
\eeq
Since $\om_{\al} > 0$, the rate is nonzero only if $\eps > 0$, i.e., if energy
conservation requires the {\it emission\/} of phonons. To perform the sum over phonon
modes, we replace the mode index $\al$ by the pair $(\bq, s)$, where
$\bq$ is the wavevector, and $s$ (= 1, 2, or 3) labels the three accoustic modes. The sum
over $\bq$ can be turned into an integral in the usual way, and we get
\beq
\Gam_{\rm pet}
    = {2\pi\by \hbar} \left( {AD \Dta' \by E_{\rm ex}} \right)^2
        {\hbar v_0 \by 8M} \sum_s \int {d^3q \by (2\pi)^3} {1\by \om_{\bq s}} 
            [ (e_{s,x} q_z + e_{s,z} q_x)^2 + (e_{s,y} q_z + e_{s,z} q_y)^2 ]
                \dta(\hbar\om_{\bq s} - \eps), 
                  \label{Gpet2}
\eeq
where $v_0$ is the volume of a unit cell. It should be remembered that $\bev_s$ depends on $\bq$.

To proceed further and obtain an order of magnitude estimate, we make the
simplifying assumption that the material is isotropic, so that for any $\bq$, there is one
longitudinal mode and two (degenerate) transverse modes, and that $\om_{\bq s}$ is either
$c_L q$ or $c_T q$, where $c_L$ and $c_T$ are the longitudinal and transverse sound velocities.
The integral and sum over $\bq$ and $s$ are then elementary, and we get
\beq
\Gam_{\rm pet}
    = {1\by 5\pi} \left( {AD \Dta' \by E_{\rm ex}} \right)^2
          \left({1\by c_T^5} + {1\by 3c_L^5} \right)
        {\eps^3 \by \rho\hbar^4},
                  \label{Gamma}
\eeq
where
\beq
\rho = M/v_0
\eeq
is the mass density of the material. The $\eps^3$ depedence of the rate (\ref{Gamma}) is chacteristic
of other rates involving phonon emission, and the $A^2 \sim S^3$ factor has the same origin as
in Ref~\cite{vhsr94}. Since $c_T \sim c_L/2$, the transverse sound term is likely to dominate
in \eno{Gamma}.

\section{The Landau-Zener-St\"uckelberg Rate}
\label{lzs}

Next, let us consider how the phonoemissive process affects the spin-flip probability if the
longitudinal field is swept in a LZS protocol. The role of $\Dta$ in \eno{p_flip} is now played by the
second-order matrix element $V^{(2)}_{\al i}$. We will only consider the limit of fast sweep, in
which case the answer may be found by a perturbative expansion in $V^{(2)}_{\al i}$. At the same time,
it should be noted that the sweep is slow on the time scale $\hbar/E_{\rm ex}$ required to establish the
underlying second-order process.

With the same notation as in \Sno{fgr}, the time-dependent Hamiltonian for the LZS protocol may be written
as
\beq
\ham_{\rm LZS} = E_i(t) \ket{i}\bra{i} + \sum_{\al} E_{\al}(t) \ket{\al}\bra{\al}
                  +\sum_{\al} \left[ V^{(2)}_{\al i} \ket{i}\bra{\al} + {\rm h.c.} \right],
\eeq
where
\bea
E_i(t) &=& \eps_{-10} + g\mu_B S {\dot H}_z t, \\
E_{\al}(t) &=& \eps_{10} + \hbar\om_{\al} - g\mu_B S {\dot H}_z t.
\eea
Hence,
\beq
\dot\eps = 2g\mu_B {\dot H}_z.
\eeq
Further, the energies $\eps_{\pm 10}$ of a single molecule are now the eigenvalues of $\ham_{\rm s}$
with $\bH = 0$. It follows that that these energies are equal, and we may take them as our reference
level:
\beq
\eps_{10} = \eps_{-10} = 0.
\eeq

In an interaction picture, the time-dependent state of the system can be written as
\beq
\ket{\psi(t)} = a(t) e^{-i\dot\eps t^2/4\hbar} \ket{i}
     + \sum_{\al} b_{\al}(t) e^{-i[\om_{\al}t - \dot\eps t^2/4\hbar]} \ket{\al}.
\eeq
Feeding this form into the Schr\"odinger equation, we obtain the equations of motion for the
amplitudes $a(t)$ and $b_{\al}(t)$:
\bea
\dot a &=& -{i\by \hbar} \sum_{\al} V^{(2)}_{i\al} b_{\al}(t) 
                    e^{-i[\om_{\al}t - \dot\eps t^2/2\hbar]},  \nnu \\
{\dot b}_{\al} &=& -{i\by \hbar} V^{(2)}_{\al i} a(t) 
                    e^{i[\om_{\al}t - \dot\eps t^2/2\hbar]}.
                               \label{eom_ab}
\eea
We wish to solve these equations with the initial conditions
\beq
|a(-\infty)| = 1, \quad b_{\al}(-\infty) = 0.
\eeq
Following Kayanuma~\cite{kaya84}, the solution may be obtained as a power series in $V^{(2)}$ by iteratively
substituting Eqns.~(\ref{eom_ab}) into one another. For us it suffices to take only the first
order answer, so we put $a(t) = 1$ in the second equation, which can then be integrated to yield
\beq
b_{\al}(t) \simeq {-i\by \hbar} V^{(2)}_{\al i}
                    \int_{-\infty}^t dt_1\, e^{i[\om_{\al}t_1 - \dot\eps t_1^2/2\hbar]}.
\eeq
Thus, ignoring a unimodular multiplicative factor,
\beq
b_{\al}(\infty) = -{i\by \hbar} V^{(2)}_{\al i} \sqrt{2\pi\hbar \by i{\dot\eps}}.
\eeq
The net spin-flip probability is 
\beq
p^{(2)}_{\rm LZS} = \sum_{\al} |b_{\al}(\infty)|^2
\eeq
The sum over phonon modes is performed in the same way as in \Sno{fgr}, except that since there is no
delta function of energy in the summand, one must cut the sum off in some way. The natural cutoff is
provided by the peak value of the bias between the $m=\pm 10$ states due to the swing in the longitudinal
field. Writing this swing as $H_{\rm ac}$, we define
\beq
E_{\rm ac} = g\mu_B S H_{\rm ac}.
\eeq
We then find
\beq
p^{(2)}_{\rm LZS} =
    {1\by 10\pi} \left( {AD \Dta' \by E_{\rm ex}} \right)^2
          \left({1\by c_T^5} + {1\by 3c_L^5} \right)
                   {E_{\rm ac}^4 \by \rho\hbar^4 \dot\eps}.
                       \label{p2lzs}
\eeq

The indirect flip probability must be added to the direct process result (\ref{pfast}). The
result can then be cast in the same form as \eno{pfast} itself, provided we replace $\Dta$
by an effective splitting $\Dta_{\rm eff}$, where
\beq
\Dta^2_{\rm eff}
   = \Dta^2 + \Dta'^2\left[
    {1\by 5\pi^2} \left( {AD \by E_{\rm ex}} \right)^2
          \left({1\by c_T^5} + {1\by 3c_L^5} \right)
                   {E_{\rm ac}^4 \by \rho\hbar^3} \right].
                       \label{Dta_eff}
\eeq

The term in square brackets provides a quick way to see the relative importance of
the phonoemisive process. If we take $H_{\rm ac} = 100$\,Oe, then
$E_{\rm ac}/\hbar = 1.8\times 10^{10}$\,sec$^{-1}$. We may take $D = 0.25$\,K, and
$E_{\rm ex} = 5$\,K as reasonable estimates. For the sound speeds and density, we may take
representative values for common organic materials. For paraffin wax, for example,
$\rho = 0.91$\,g/cm$^3$, $c_L = 1.94 \times 10^5$\,cm/s, and we have not been able
to find a value for $c_T$. For polyethylene, 
$\rho = 0.90$\,g/cm$^3$, $c_L = 1.95 \times 10^5$\,cm/s, and $c_T = 0.54\times 10^5$\,cm/s,
and for polystyrene,
$\rho = 1.06$\,g/cm$^3$, $c_L = 2.35 \times 10^5$\,cm/s, and $c_T = 1.12\times 10^5$\,cm/s.
Hence we take $\rho = 1$\,g/cm$^3$, and $c_T = 10^5$\,cm/s. The term in square brackets is
then of order $10^{-12}$, which is disappointingly small. Even if we take
$\Dta' \simeq 100 \Dta$ as is appropriate if the intermediate state is $m=\pm 8$ in \Fe8, the
contribution of the phonoemissive process is  negligible. Nevertheless, the process remains
the simplest imaginable candidate for the magnetization relaxation mechanism. To this author's
mind, it remains an urgent problem to resolve this puzzle.

\begin{acknowledgement}
This work was begun with support from the NSF via grant number
DMR-0202165.
\end{acknowledgement}

\end{document}